# Phonon States in NbTe$_4$ and TaTe$_4$ Quasi-One-Dimensional van der Waals Crystals

Zahra Ebrahim Nataj,[1,2] Fariborz Kargar,[1,2*] Sergiy Krylyuk,[3] Topojit Debnath,[2] Maedeh Taheri,[1,2] Subhajit Ghosh,[1,2] Huairuo Zhang,[3,4] Albert V. Davydov,[3] Roger K. Lake,[2] and Alexander A. Balandin[1,2*]

[1]Department of Materials Science and Engineering, University of California, Los Angeles, California 90095 USA

[2]Department of Electrical and Computer Engineering, University of California, Riverside, California 92521 USA

[3]Materials Science and Engineering Division, National Institute of Standards and Technology, Gaithersburg, Maryland 20899 USA

[4]Theiss Research, Inc., La Jolla, California 92037 USA

---

[*] Corresponding authors: f.kargar@ucla.edu (F.K.); balandin@seas.ucla.edu (A.A.B); website: https://balandin-group.ucla.edu/

# Abstract

We report the results of polarization-dependent Raman spectroscopy of phonon states in single-crystalline quasi-one-dimensional NbTe$_4$ and TaTe$_4$ van der Waals materials. The measurements were conducted in the wide temperature range from 80 K to 560 K. Our results show that although both materials have identical crystal structures and symmetries, there is a drastic difference in the intensity of their Raman spectra. While TaTe$_4$ exhibits well-defined peaks through the examined frequency and temperature ranges, NbTe$_4$ reveals extremely weak Raman signatures. The measured spectral positions of the phonon peaks agree with the phonon band structure calculated using the density-functional theory. We offer possible reasons for the intensity differences between the two van der Waals materials. Our results provide insights into the phonon properties of NbTe$_4$ and TaTe$_4$ van der Waals materials and indicate the potential of Raman spectroscopy for studying charge-density-wave quantum condensate phases.

**I. INTRODUCTION**

In recent years, there has been a growing interest in one-dimensional (1D) van der Waals (vdW) materials, which reveal interesting phenomena, such as nontrivial band topology, superconductivity, magnetism, and charge-density-wave (CDW) formation.[1–5] The CDW quantum condensate phase, often observed in quasi-1D materials, is a periodic modulation of the electron charge density that is accompanied by a periodic distortion of the underlying crystal lattice.[6–8] Transition-metal tri-chalcogenides (MX$_3$, M: transition metal, X: chalcogen), transition metal tri-halides, and transition metal tetra-chalcogenides (MX$_4$) are examples of compounds that encompass a variety of 1D structures.[2,9–11] Among them, NbTe$_4$ and TaTe$_4$ attracted particular attention owing to their CDW phases.[12–19] A recent study suggested that the charge order modulation in NbTe$_4$ can transform it into a Weyl semimetal.[12] In the case of TaTe$_4$, it was proposed that the charge order drives the system in a topological state with double Dirac fermions.[20]

In 1D van der Waals materials, atoms form one-dimensional chains through strong covalent bonds. These chain structures are interconnected through weaker covalent, ionic, or van der Waals bonds along other crystallographic directions.[1,2] Generally, 1D materials can be classified into "true-1D" and "quasi-1D" systems, depending on the relative strength of the atomic bonds along different crystallographic directions compared to the chain direction.[1,2] In "true-1D" materials, atoms form strong covalent bonds along the atomic chains, accompanied by weak van der Waals bonds along all other directions.[1,9,10,21] Quasi-1D crystals also feature strong covalent bonds along the atomic chains but exhibit weaker covalent or ionic bonds along other directions. As a consequence of the varying bond strengths, true-1D materials, like Nb$_2$Se$_9$ and MoI$_3$, adopt needle-like structures with extremely high geometrical aspect ratios, whereas quasi-1D crystals, such as ZrTe$_3$, display ribbon-like structures after exfoliation.[1,9,10,21,22] Adopting this classification, NbTe$_4$ and TaTe$_4$, both fall into the category of quasi-1D structures owing to their unique tetragonal crystal structure.[11] One may also say that NbTe$_4$ and TaTe$_4$ belong to the intermediate 1D/2D class of van der Waals materials.[23]

Raman spectroscopy is the conventional way to probe the energy of optical phonons; it has been extensively used as a non-destructive tool to identify the material, assess its quality, and provide deeper insights into the material's properties.[9,24–29] Phonons and their interaction with electrons play an important role in the formation of CDW phases. The signatures of CDW phases are observed in the temperature-dependent Raman spectra in various ways, such as the appearance or disappearance of certain Raman features, changes in their spectral position, full-width-half-maximum, or intensity.[30–32] For example, in 1*T*-TaS$_2$, a quasi-2D CDW material, the Raman spectra evolve from a single broad peak in the low-frequency regions to many distinguishable Raman peaks in the low and high-frequency ranges as the material cools down from 350 K to 80 K. The appearance of additional peaks at the low-frequency region is due to the formation of the periodic superstructure, and the corresponding phonon zone-folding in the commensurate CDW phase.[32–35] In other materials, one can observe CDW amplitude modes or additional phonon peaks emerging due to the relaxation of the Raman selection rules as a result of crystal lattice distortion associated with the CDW formation.[36] In order to assess the possibility of using Raman spectroscopy as a tool to monitor strongly correlated phenomena in van der Waals materials, systematic spectroscopy data is required. The information about phonon states in NbTe$_4$ and TaTe$_4$ is either unavailable or scarce. We are only aware of one study, which reported the Raman spectrum of TaTe$_4$.[15] The additional difficulties with the interpretation of Raman spectrum features in van der Waals materials, typically synthesized *via* the vapor transport methods, are associated with possible deviations from the stoichiometry, unintentional doping, and defects that may affect the Raman selection rules.[37,38] In this paper, we report the results of the combined experimental and computational study of phonon states in quasi-1D NbTe$_4$ and TaTe$_4$ materials and discuss possible physical mechanisms behind the drastic difference in the intensity of the Raman spectra.

**II. MATERIALS SYNTHESIS AND CHARACTERIZATION**

The materials for this study were grown using chemical vapor transport (CVT) method. In the case of NbTe$_4$ crystals, the stoichiometric amounts of Nb and Te, with a total weight of 2.6 g as well as 90 mg of I$_2$ as the transport agent were sealed in a quartz ampoule under vacuum. The ampoule was placed in the three-zone furnace and heated at ~1 °C/min until the temperature of the charge, *i.e.*, where the precursors are sitting, and the temperature of the growth zones of the ampoule, *i.e.*, where the crystals grow, reached 800 ºC and 700 ºC, respectively. After 5 d of growth, the furnace was shut off and cooled at ambient temperature. The TaTe$_4$ crystal was grown in a similar manner using stoichiometric amounts of Ta and Te powders and TeCl$_4$ as the transport agent. The temperature at the charge and growth zones was 550 °C and 530 °C, respectively, and the growth duration was 5 d.

Figure 1 (a, b) presents the crystal structure of the NbTe$_4$ and TaTe$_4$ along different crystallographic directions. Both crystals have a tetragonal crystal lattice structure belonging to the space group of P4/mcc (124).[12,15,39] The Te atoms (blue spheres) are arranged in chains of slightly distorted square antiprisms along the *c*-axis, with Nb or Ta atoms (red spheres) sitting at the center of these square antiprisms. The crystal structure and symmetry of the compounds were further verified using high-angle annular dark-field scanning transmission electron microscopy (HAADF-STEM) imaging (vide infra) and X-ray diffraction (XRD). The STEM image of NbTe$_4$ and TaTe$_4$ along the [010] zone axis is presented in Figure 1 (c, d), respectively. The obtained STEM images for both samples match with the crystal structure schematic presented in panel (b). Figure 1 (e) exhibits the diffraction patterns of NbTe$_4$ (black line) and TaTe$_4$ (red line) with selected *hkl* indices. Lattice parameters of $a = b = 6.5029(2)$ Å, $c = 6.833(9)$ Å for NbTe$_4$ and $a = b = 6.5148(3)$ Å, $c = 6.811(1)$ Å for TaTe$_4$ were obtained which agree well with the data previously reported by others.[39,40] The numbers in the parentheses represent the standard deviation values. Based on the STEM images and the diffraction patterns, P4/mcc (124) space group symmetry was confirmed for both crystal structures. Figure 1 (f, g) shows scanning electron microscopy (SEM) images of representative NbTe$_4$ and TaTe$_4$ samples mechanically exfoliated on Si/SiO$_2$ substrates, respectively. After exfoliation, the samples appear in nanoribbon structures owing to their quasi-1D nature. More SEM images of exfoliated crystals for both NbTe$_4$ and TaTe$_4$ are presented in Supplementary Figure S1.

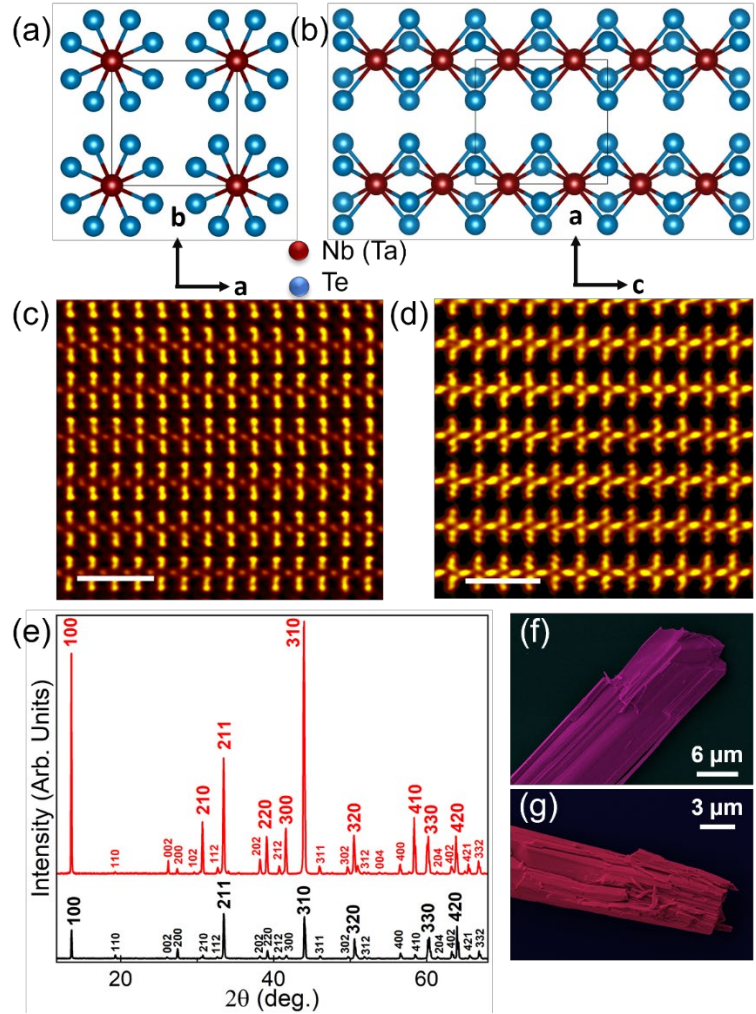

[Figure 1: Crystal structure of NbTe$_4$ and TaTe$_4$ shown along the (a) *c*, and (b) *b* crystallographic directions. Red and blue spheres denote Nb (Ta) and Te atoms, respectively. (c, d) Atomically-resolved HAADF-STEM images of (c) NbTe$_4$ and (d) TaTe$_4$ crystals along the [010] zone axis. Note the excellent agreement of the STEM images with the crystal structure schematic presented in panel (b). The scale bars exhibit 1 nm. (e) XRD patterns of NbTe$_4$ (bottom, black line) and TaTe$_4$ (top, red line) crystals with selected *hkl* indices. (f, g) Representative SEM images of (f) NbTe$_4$ and (g) TaTe$_4$ samples mechanically exfoliated on Si/SiO$_2$ substrates.]

### III. RAMAN SPECTRA AND PHONON BANDSTRUCTURE

We conducted polarization-dependent Raman spectroscopy over a wide temperature range to investigate the phonon properties of NbTe$_4$ and TaTe$_4$. All experiments were carried out in the backscattering configuration under a 488 nm laser excitation wavelength. The laser power on samples was adjusted at 450 µW to prevent any laser-induced local heating effects. Figure 2 (a,

b) presents the results of Raman measurements of both crystals at room temperature (RT) when the polarization of the incident light, $e^i$, is parallel with and transverse to the atomic chains, *i.e.*, the crystal's *c*-axis. The polarization of the scattered light, $e^s$, was not analyzed. In both crystals, the Raman spectra strongly depend on the polarization of the incident light, a characteristic of quasi-1D materials. The NbTe$_4$ and TaTe$_4$ crystals each have 10 atoms in the primitive cell with a total of 27 optical phonon branches. Since the Raman spectral features of NbTe$_4$ are rather weak, one can use a comparative approach to identify the Raman peaks of NbTe$_4$ based on those observed for TaTe$_4$. Taking this approximate approach, we determined the spectral position of the Raman peaks of both crystals by fitting individual Lorentzian functions to the accumulated experimental data. The vibrational symmetry of each phonon mode was also determined by the theoretical calculations. An important observation from Figure 2 is that there is a strong difference in the intensity of Raman spectra obtained for NbTe$_4$ and TaTe$_4$ under similar experimental conditions. One can clearly see that the intensity of Raman features of NbTe$_4$ is significantly lower than that for TaTe$_4$ in both polarization configurations. We address possible reasons for such a difference in the Discussion Section.

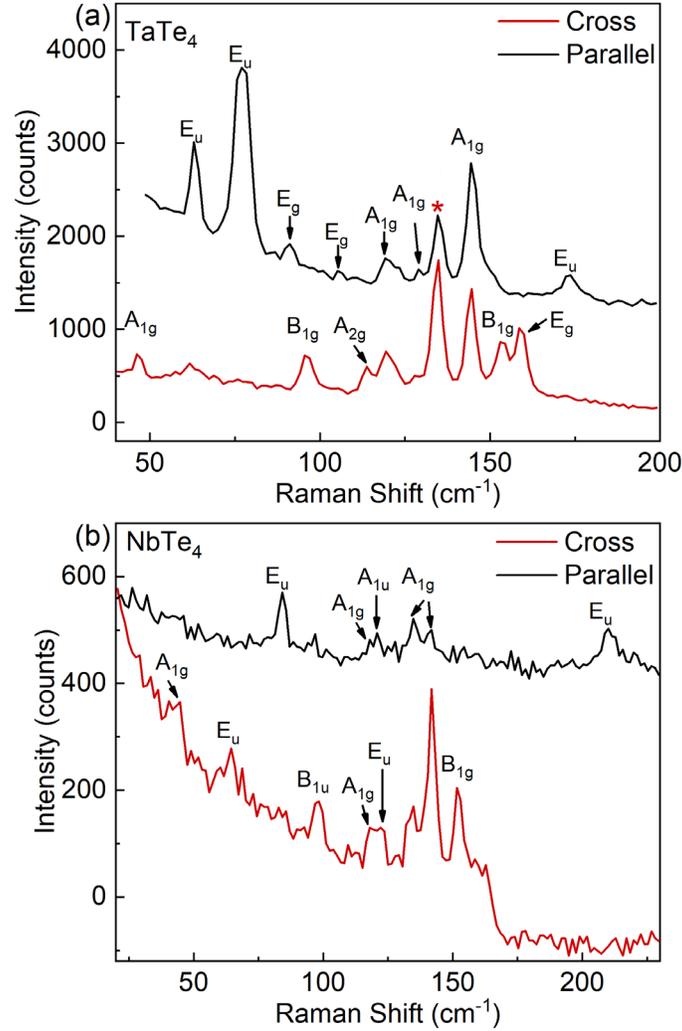

[Figure 2: Raman spectra of (a) TaTe$_4$ and (b) NbTe$_4$ at room temperature obtained at parallel, $e^i \parallel c$, and cross, $e^s \perp c$, polarization configurations.]

In order to interpret Raman spectral features, we calculated the phonon dispersion of both compounds using the generalized gradient approximation (GGA),[41] exchange-correlation functional of Perdew, Burke, and Ernzerhof (PBE),[42,43] and projected augmented wave (PAW) method as implemented in the Vienna *ab initio* Simulation Package (VASP).[44] A cutoff of 520 eV was chosen for the plane wave basis set. The relaxation was done with a conjugate gradient algorithm until the difference in total energy and forces on the atoms were smaller than $10^{-9}$ eV and $10^{-8}$ Å$^{-1}$, respectively. The Brillouin zone was sampled with a 12 x 12 x 12 k-point grid and the vdW interactions were accounted using the DFT-D3 method proposed by Grimme.[45] The phonon dispersion was calculated using the finite displacement supercell method as implemented on Phonopy[46,47] with a grid of 6 x 6 x 6 k-points. The phonon band structure of TaTe₄ and NbTe₄ are shown in Figures 3 (a, b), respectively. The spectral position of the Raman peaks is shown with green spheres and red stars for the $e^i||c$ and $e^i \perp c$ polarization configurations, respectively. As one can see, there is a satisfactory agreement between the experimental Raman data points and the calculated phonon energies.

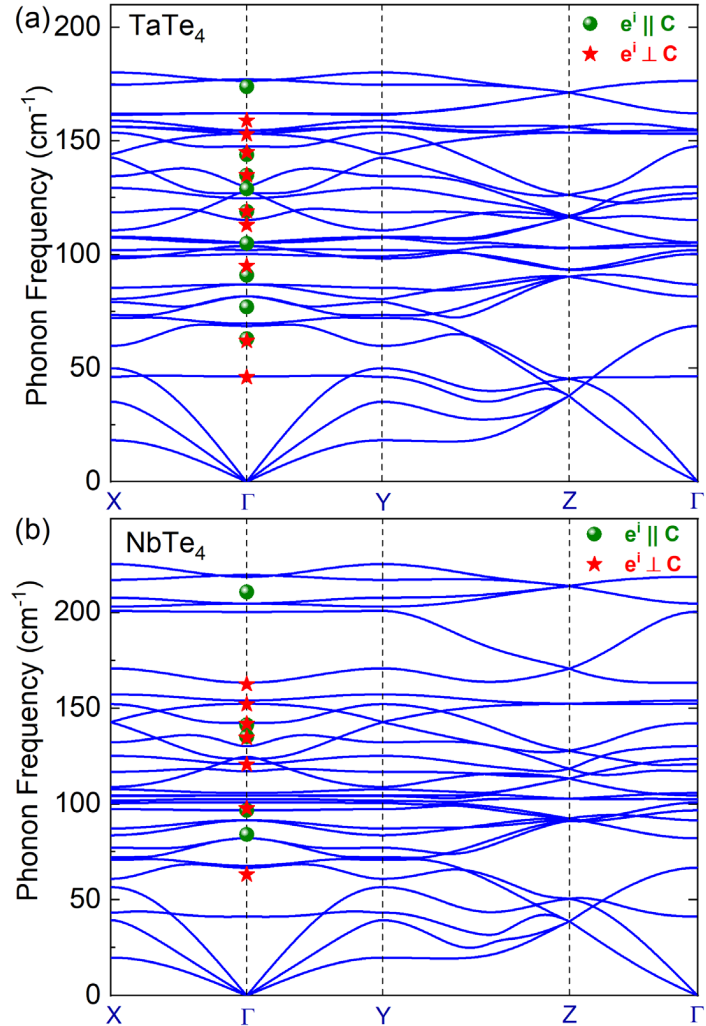

[Figure 3: Calculated phonon dispersion for (a) TaTe₄ and (b) NbTe₄. The green and red symbols are the experimental data obtained in parallel and cross-polarization for both crystals, respectively.]

Figure 4 shows the vibration profile of selected phonon modes with symmetries around the inversion point along with their respective calculated frequencies for both TaTe₄ and NbTe₄. The atomic displacement and frequencies of all optical and acoustic phonon modes are listed in Supplementary Table S1. The vibrational symmetry of atoms determines the forbidden or permitted Raman modes, *i.e.*, the modes that appear in the experimental Raman spectra out of the entire optical phonon modes of the material. In an undistorted P4/mcc structure and in the backscattering geometry, $A_{1g}$, $B_{1g}$, and $E_g$ modes are Raman active while $E_u$, $B_{2g}$ and $B_{1u}$ modes are forbidden. However, as seen in Figure 2 (a, b), some peaks assigned as $E_u$ and $B_{1u}$ symmetries that their energies match closely with the calculated phonon energies presented in Figure 3 (a, b). We will address possible reasons of appearance of these modes in the Raman spectra in the Discussion section.

|  | $A_{2g}$ | $E_g$ | $E_g$ | $B_{1g}$ | $E_g$ |
|---|---|---|---|---|---|
|  | 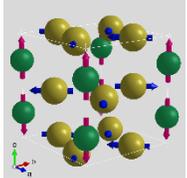 | 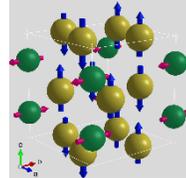 | 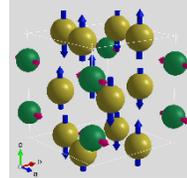 | 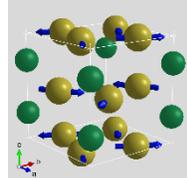 | 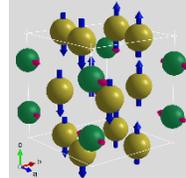 |
| NbTe₄ | 41 cm⁻¹ | 92 cm⁻¹ | 92 cm⁻¹ | 102 cm⁻¹ | 104 cm⁻¹ |
| TaTe₄ | 44 cm⁻¹ | 87 cm⁻¹ | 87 cm⁻¹ | 101 cm⁻¹ | 105 cm⁻¹ |
|  | $E_g$ | $B_{2g}$ | $A_{2g}$ | $A_{1g}$ | $A_{1g}$ |
|  | 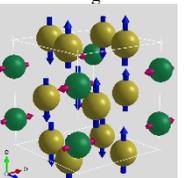 | 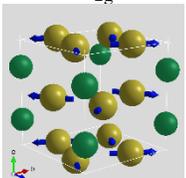 | 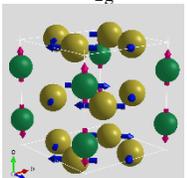 | 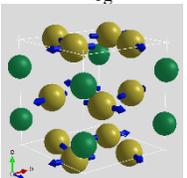 | 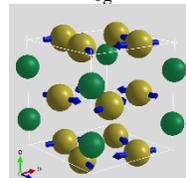 |
| NbTe₄ | 104 cm⁻¹ | 106 cm⁻¹ | 117 cm⁻¹ | 130 cm⁻¹ | 142 cm⁻¹ |
| TaTe₄ | 105 cm⁻¹ | 105 cm⁻¹ | 115 cm⁻¹ | 130 cm⁻¹ | 147 cm⁻¹ |
|  | $B_{2g}$ | $B_{1g}$ | $A_{2g}$ | $E_g$ | $E_g$ |
|  | 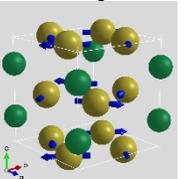 | 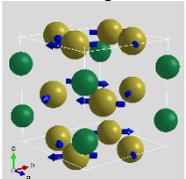 | 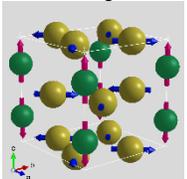 | 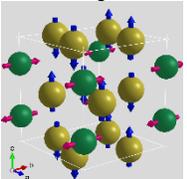 | 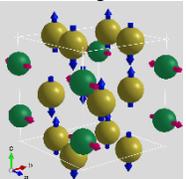 |
| NbTe₄ | 153 cm⁻¹ | 154 cm⁻¹ | 163 cm⁻¹ | 205 cm⁻¹ | 205 cm⁻¹ |
| TaTe₄ | 153 cm⁻¹ | 154 cm⁻¹ | 154 cm⁻¹ | 162 cm⁻¹ | 162 cm⁻¹ |

[Figure 4: Vibrational symmetry of phonon modes for both NbTe₄ and TaTe₄ along with their respective mode frequencies. The green and yellow spheres represent Nb (Ta) and Te atoms, respectively.]

The temperature-dependent Raman spectra can be used for the identification of possible phase transitions, irreversible structural modifications, or extraction of Raman temperature coefficients, *i.e.*, the frequency change of each phonon mode as a function of temperature. The data on the temperature coefficient provides insights into the mode-specific phonon anharmonicity and lattice expansion.[48–50] It can be also used for thermal conductivity measurements using the Raman optothermal technique.[29,51] We conducted temperature-dependent Raman experiments on exfoliated NbTe₄ samples in the temperature range of 80 K to 300 K. The results for parallel and cross-polarization light scattering configurations with temperature increments of 10 K are presented in Figures 5 (a, b), respectively. The spectra at different temperatures are translated vertically for clarification. The spectral position, $\omega$, of the intense peaks in both polarization configurations are plotted as a function of temperature and shown in Figure 5 (c, d). In both panels, the black lines are the linear regression fittings, $\omega(t) = \chi T + \omega_0$, over the experimental data, in which $\omega_0$ is the frequency of the Raman peak at absolute zero and $\chi$ is the Raman first-order temperature coefficient. As seen, the frequency of all selected peaks decreases linearly with temperature, with some modes exhibiting stronger phonon softening. The phonons redshift with the temperature rise due to the crystal lattice expansion and phonon anharmonic effects, as expected.[48] No strong or abrupt changes in the peak spectral positions were observed in the temperature interval of 150 K ≤ $T$ ≤ 200 K, the reported range in which NbTe₄ undergoes a CDW transition.[11,12,52,53]

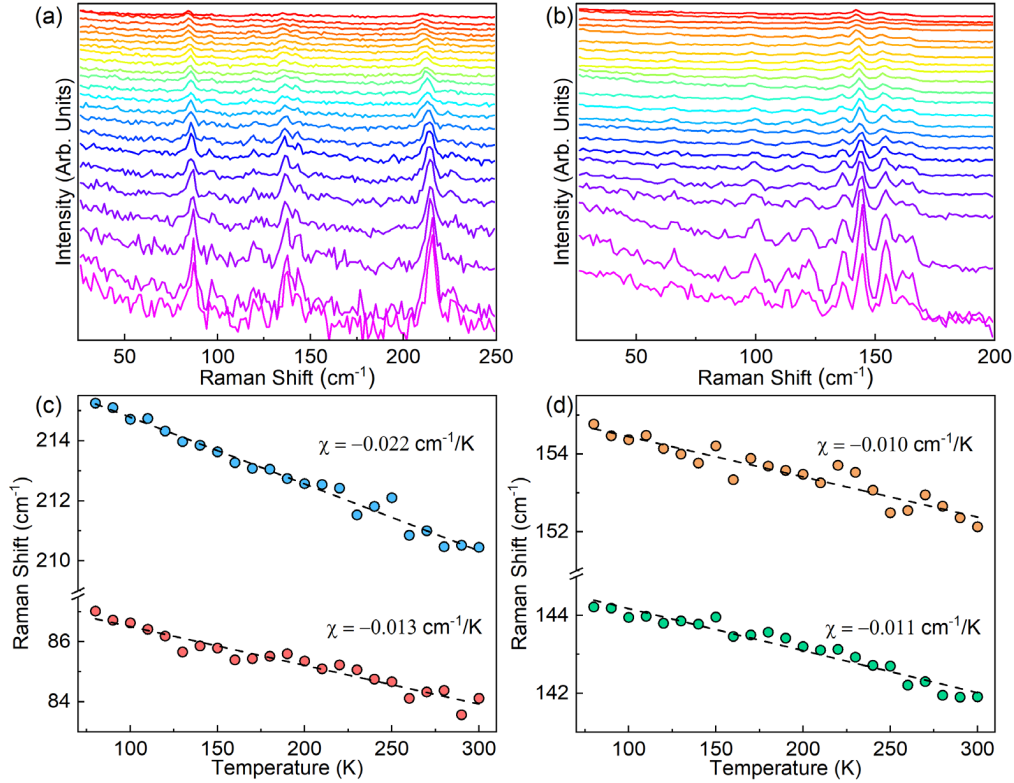

[Figure 5: Temperature-dependent Raman spectra of NbTe$_4$, ranging from 80 K (violet) to 300 K (red), displayed in a color gradient from bottom to top for (a) parallel, and (b) cross-polarization scattering configurations. The spectra were taken with temperature increments of 10 K. (c, d) The spectral position of selected peaks as a function of temperature. The dashed lines are linear regression over the experimental data with $\chi$ denoting the Raman temperature coefficient.]

We also accumulated Raman spectra of TaTe$_4$ in the temperature range of 300 K ≤ T ≤ 560 K to investigate its CDW transition reported at ~450 K. The results are presented in Figure 6 (a) with temperature steps of 10 K. As the temperature increases, the two peaks at ~113 cm$^{-1}$ and ~120 cm$^{-1}$ at RT eventually merge and form an intense broad peak located at 118 cm$^{-1}$ at 440 K. With further temperature rise and at 460 K, another broad peak appears at ~420 cm$^{-1}$. To identify the nature of changes observed in the Raman spectra, the same sample was cooled down again and another Raman measurement was conducted at RT. The Raman spectra for a freshly exfoliated sample at RT, at 570 K, and after cooling back to RT are shown in Figure 6 (b). At 570 K, two sharp peaks at 108 cm$^{-1}$ and 136 cm$^{-1}$ are formed that are likely associated with the oxidation of Te atoms or Te-metalloid-like chains.[54,55] Also note that the peak 420 cm$^{-1}$ does not disappear after cooling the sample back to RT, indicating that the process is irreversible. The latter indicates oxidation or permanent structural change in TaTe$_4$ at elevated temperatures rather than the reported high-temperature CDW transition.[11,15,56] A recent study reported the Raman signatures of TaTe$_4$ under vacuum conditions at varying temperatures and attributed changes in the intensity of a peak at 58 cm$^{-1}$ to a possible CDW transition.[15] We note that our DFT calculations for TaTe$_4$ do not reveal any phonon bands in the vicinity of the Brillouin zone center at 58 cm$^{-1}$. The reported Raman data in Ref [15] resembles that of TaTe$_2$, which exhibits a sharp peak at 58 cm$^{-1}$.[24] In Figure 6 (c, d) we present the spectral position of some well-defined Raman peaks as a function of temperature. The data is analyzed for the temperature range of RT to 440 K where the sample has not undergone structural changes. Note that the peak at ~46 cm$^{-1}$ has low intensity which induces significant uncertainty in determining its accurate spectral position and causes deviations from linear regression fitting. The data was fitted using the linear regression, $\omega(t) = \chi T + \omega_0$. All phonon modes exhibit softening with temperature rise as expected. No anomalies in the peak position were observed in the examined temperature range. While we did not observe clear signatures of the CDW phase transitions, *e.g.*, appearance of folded phonon modes in commensurate CDW phase or disappearance of peaks due to the loss of translation symmetry in incommensurate CDW phases, below we argue that the drastic difference in the Raman intensity of NbTe$_4$ and TaTe$_4$ quasi-1D van der Waals crystals may be related to CDW phenomena.

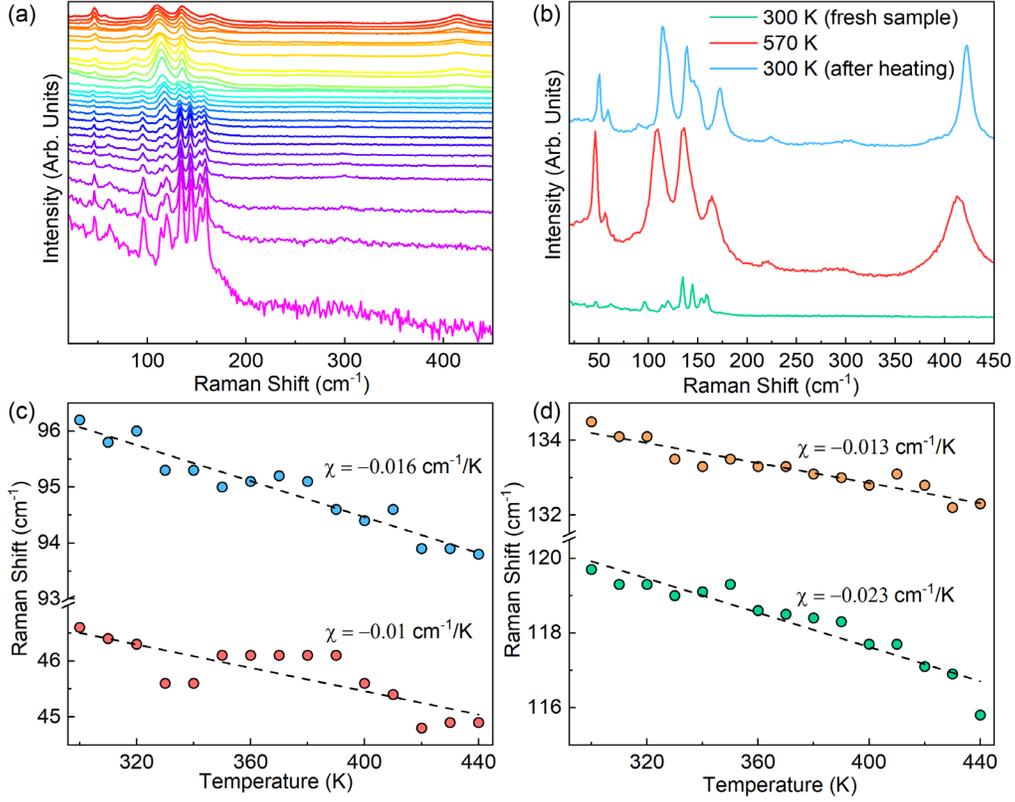

[Figure 6: Temperature-dependent Raman spectra of TaTe₄, ranging from 300 K (violet) to 570 K (red), displayed in a color gradient from bottom to top. The spectra were taken with temperature increments of 10 K. (b) Raman spectra of TaTe₄ during a heating-cooling cycle. The green, red, and blue lines present the data for a freshly exfoliated sample at RT, at 570 K, and the same sample cooled down to RT, respectively. (c, d) The spectral position of selected peaks in (a) as a function of temperature. The dashed lines are linear regression over the experimental data with $\chi$.denoting the Raman temperature coefficient.]

## IV. DISCUSSION

Below 50 K, NbTe₄ is in a fully commensurate CDW (C–CDW) state with its underlying lattice. With the temperature increase, the C–CDW phase evolves into the discommensurate CDW state (D–CDW) in the temperature range between ~50 K to 60 K. With further increasing the temperature, the material undergoes another transition between ~150 K to 200 K into the incommensurate CDW (IC–CDW) phase.[53] TaTe₄ exhibits drastically different CDW characteristics. The material is commensurately modulated even at room temperature. Upon heating, the crystal exhibits a commensurate-to-commensurate phase transition at ~450 K and subsequently, a C–CDW to IC–CDW phase at 550 K.[16] These transitions have been reported previously for both tetratellurides using a variety of other experimental techniques such as X-ray diffraction (XRD), selected area electron diffraction (SAED), satellite dark field electron microscopy (SDFEM), and scanning tunneling microscopy (STM).[11,12,57–63,13,15,16,19,20,52,53,56] One should note that the CDW transitions are accompanied by distortions in the lattice and can affect the dynamics of the vibrational symmetry and Raman signatures of the material. Let us now consider the multitude of observed Raman peaks of different symmetries and intensity disparity for NbTe₄ and TaTe₄ quasi-1D van der Waals crystals (see Figure 2 (a, b)).

The differential scattering cross section (DSCS) described as $\frac{d\sigma}{d\Omega} \sim \left| e_k^s \, x_{kl} \, e_l^i \right|^2$, can be used to determine the forbidden and allowed Raman modes. Here, $e_l^i$ and $e_k^s$ are the elements of the polarization vector of the incident and scattered light and $x_{kl}$ are the elements of the Raman tensor of the respective vibrational symmetry. Note that in our experiments, the polarization of the incident light was aligned either parallel or transverse to the atomic chains, *i.e.*, crystallographic c-axis, while the polarization of the scattered light was not analyzed. The DSCS of each mode in the backscattering geometry and for the P4/mcc crystal structure is calculated and listed in Table I. The DSCS of modes with $A_{1g}$ and $E_g$ symmetries is nonzero in both cases of incident light polarizations, and therefore, they are observable in Raman spectra when $e^i || c$ or $e^i \perp c$. The DSCS of $B_{1g}$ mode is zero when $e^i || c$ and nonzero when $e^i \perp c$. The DSCS of the $B_{2g}$ vibrational modes are zero, thus they are forbidden in both incident light polarizations. All modes with Eu or Au symmetries in an *undistorted* P4/mcc crystal symmetry are Raman inactive. The fact that these modes are present in the spectra shown in Figure 2 (a, b) might indicate the relaxation of Raman selection rules due to even small lattice distortions caused by CDW effects and a consequent breaking of inversion symmetry. The sharp peak indicated with an asterisk in Figure 2 (a) for TaTe₄ is not associated with any theoretical phonon bands and its origin is unclear to us. Similar anomalies such as observation of Raman-inactive modes or peaks without any corresponding calculated values have been reported for other CDW materials such as GdTe₃.[36] Thus, while there are no abrupt changes in the Raman spectra of these materials at the phase transition points, the appearance of the forbidden peaks may indicate lattice distortion associated with CDW phases. The loss of translation symmetry due to boundaries and defects may also result in the

relaxation of the Raman selection rules. However, the cross-sectional dimensions of our exfoliated crystals are rather large preventing the phonon spectrum changes. The material characterization data suggest a high-quality material with low defects.

Table I: The vibrational symmetry and forbidden (*f*) and permitted (*p*) Raman modes

| | Parallel Polarization | | Cross Polarization | |
|---|---|---|---|---|
| $A_{1g}$ | $\left\| [0\ 1\ 1] \begin{bmatrix} a & 0 & 0 \\ 0 & a & 0 \\ 0 & 0 & b \end{bmatrix} \begin{bmatrix} 0 \\ 0 \\ 1 \end{bmatrix} \right\|^2 = b^2$ | p | $\left\| [0\ 1\ 1] \begin{bmatrix} a & 0 & 0 \\ 0 & a & 0 \\ 0 & 0 & b \end{bmatrix} \begin{bmatrix} 0 \\ 1 \\ 0 \end{bmatrix} \right\|^2 = a^2$ | p |
| $B_{1g}$ | $\left\| [0\ 1\ 1] \begin{bmatrix} c & 0 & 0 \\ 0 & -c & 0 \\ 0 & 0 & 0 \end{bmatrix} \begin{bmatrix} 0 \\ 0 \\ 1 \end{bmatrix} \right\|^2 = 0$ | f | $\left\| [0\ 1\ 1] \begin{bmatrix} c & 0 & 0 \\ 0 & -c & 0 \\ 0 & 0 & 0 \end{bmatrix} \begin{bmatrix} 0 \\ 1 \\ 0 \end{bmatrix} \right\|^2 = c^2$ | p |
| $B_{2g}$ | $\left\| [0\ 1\ 1] \begin{bmatrix} 0 & d & 0 \\ d & 0 & 0 \\ 0 & 0 & 0 \end{bmatrix} \begin{bmatrix} 0 \\ 0 \\ 1 \end{bmatrix} \right\|^2 = 0$ | f | $\left\| [0\ 1\ 1] \begin{bmatrix} 0 & d & 0 \\ d & 0 & 0 \\ 0 & 0 & 0 \end{bmatrix} \begin{bmatrix} 0 \\ 1 \\ 0 \end{bmatrix} \right\|^2 = 0$ | f |
| $E_g$ | $\left\| [0\ 1\ 1] \begin{bmatrix} 0 & 0 & 0 \\ 0 & 0 & e \\ 0 & e & 0 \end{bmatrix} \begin{bmatrix} 0 \\ 0 \\ 1 \end{bmatrix} \right\|^2 = e^2$ | p | $\left\| [0\ 1\ 1] \begin{bmatrix} 0 & 0 & 0 \\ 0 & 0 & e \\ 0 & e & 0 \end{bmatrix} \begin{bmatrix} 0 \\ 1 \\ 0 \end{bmatrix} \right\|^2 = e^2$ | P |
| $E_g$ | $\left\| [0\ 1\ 1] \begin{bmatrix} 0 & 0 & -e \\ 0 & 0 & 0 \\ -e & 0 & 0 \end{bmatrix} \begin{bmatrix} 0 \\ 0 \\ 1 \end{bmatrix} \right\|^2 = 0$ | f | $\left\| [0\ 1\ 1] \begin{bmatrix} 0 & 0 & -e \\ 0 & 0 & 0 \\ -e & 0 & 0 \end{bmatrix} \begin{bmatrix} 0 \\ 1 \\ 0 \end{bmatrix} \right\|^2 = 0$ | f |

To interpret the drastic difference between the Raman intensities of NbTe₄ and TaTe₄ quasi-1D van der Waals crystals, we recall that the Raman scattering efficiency, $dS/d\Omega$, which is related to the Raman susceptibility, $a$, can be calculated from the following equation:[64]

$$\frac{dS}{d\Omega} = \left(\frac{\omega_l(\omega_l - \omega_{ph})^3}{c^4}\right) \frac{\hbar[N(\omega_{ph}) + 1]}{2V_c \mu \omega_{ph}} a^2. \qquad (1)$$

In this equation, $\omega_l$ is the frequency of the excitation laser, and $\omega_{ph}$ and $N(\omega_{ph})$ are the angular frequency and the Bose-Einstein distribution of the specific phonon mode, respectively. $V_c$ and $\mu$ are the volume and reduced mass of the unit cell and $c$ is the speed of light. The measurable quantity in Raman scattering, *i.e.*, the number of scattered photons $R_s$ within a solid collection angle $\Delta\Omega$ is related to the Raman scattering efficiency and optical properties of the materials as follows:[64]

$$R_s = \frac{8}{c^4 V_c \mu} \frac{\omega_s^3}{\omega_{ph}} \frac{N(\omega_{ph})+1}{(n+1)^4} P_l \Delta\Omega L a^2. \qquad (2)$$

In this equation, $n$ is the refractive index of the material at the laser excitation wavelength, $L$ is the scattering length, and $P_l$ is the incoming laser power. Nb atoms are about two times lighter than the Ta atoms, and thus, the $\mu$ for NbTe₄ is smaller than that of TaTe₄. The $V_c$ of both crystals are ~304 Å³. Therefore, the drastic difference between the intensity of Raman features in these two crystals can be attributed either to their refractive index, $n$, and/or Raman susceptibility, $a$. Given that the reported electrical resistivity of both materials is nearly the same at RT,[53] it can be deduced that the refractive indices of the two materials are within a similar order of magnitude. Thus, Raman susceptibility plays a key role in the intensity difference between two crystals. The Raman susceptibility depends on the material's polarizability and cannot be measured directly. At RT, TaTe₄ is still in the commensurate CDW phase whereas NbTe₄ has already transformed into the incommensurate CDW state with stronger metallic behavior. The polarizability of a metal-like medium is weak, which results in weaker Raman features.[65] Based on the above considerations, we argue that even though we cannot clearly observe the CDW phase transitions in NbTe₄ and TaTe₄, the differences in Raman intensity between these two materials are associated with CDW quantum condensate phases.

## V. CONCLUSIONS

We investigated the optical phonon properties of NbTe₄ and TaTe₄, two quasi-1D vdW materials, using temperature- and polarization-dependent Raman spectroscopy. It was found that TaTe₄ reveals well-defined Raman peaks while NbTe₄ exhibits extremely weak Raman signatures. The latter is likely explained by the fact that NbTe$_4$ is in an incommensurate CDW phase in the examined temperature range. This phase is characterized by a more metallic behavior and thus, weaker polarizability. The temperature-dependent Raman spectra of TaTe$_4$ show irreversible peaks emerging at ~460 K upon heating, indicating the possibility of oxidation or permanent crystal reconfiguration at higher temperatures. The spectral position of the Raman peaks was compared to the phonon band structure calculations obtained from the density-functional theory calculations. A good agreement between the theory and experiment was achieved. Our results shed light on the phononic properties of quasi-1D NbTe$_4$ and TaTe$_4$ and add to the knowledge of the properties of these important topological van der Waals materials.

**Acknowledgments**

A.A.B. acknowledges the support of the National Science Foundation (NSF) program Designing Materials to Revolutionize and Engineer our Future (DMREF) *via* a project DMR-1921958 entitled Collaborative Research: Data-Driven Discovery of Synthesis Pathways and Distinguishing Electronic Phenomena of 1D van der Waals Bonded Solids. A.A.B. was also supported by the Vannevar Bush Faculty Fellowship from the Office of Secretary of Defense (OSD) under the Office of Naval Research (ONR) contract N00014-21-1-2947 on One-Dimensional Quantum Materials. DFT calculations were supported by the ONR award and used STAMPEDE2 at TACC through allocation DMR130081 from the Advanced Cyberinfrastructure Coordination Ecosystem: Services & Support (ACCESS) program, which is supported by National Science Foundation grants #2138259, #2138286, #2138307, #2137603, and #2138296. H.Z. acknowledges support from the U.S. Department of Commerce, NIST under financial assistance awards 70NANB22H101 and 70NANB22H075. S.K. and A.V.D. acknowledge support from the Material Genome Initiative funding allocated to NIST.

**Disclaimer**

**Contributions**

F.K. and A.A.B. conceived the idea, coordinated the project, contributed to experimental data analysis, and led the manuscript preparation. Z.E.N. conducted sample characterization, Raman spectroscopy measurements, and contributed to the data analysis. S.K. synthesized bulk crystals by CVT. T.D. conducted *ab initio* simulations of the phonon dispersion; M.T. and S.G. contributed to sample characterization; H.Z. conducted the TEM characterization; A.V.D. supervised material growth, characterization, and contributed to data analysis. R.K.L. contributed to the data analysis and supervised the simulations. All authors contributed to the manuscript preparation and editing.

**References**

[1]     J. Lee, B. J. Kim, Y. K. Chung, W. G. Lee, I. J. Choi, S. Chae, S. Oh, J. M. Kim, J. Y. Choi, J. Huh, *J. Raman Spectrosc.* **2020**, *51*, 1100.

[2]     A. A. Balandin, F. Kargar, T. T. Salguero, R. K. Lake, *Mater. Today* **2022**, *55*, 74.

[3]     C. W. Chen, J. Choe, E. Morosan, *Reports Prog. Phys.* **2016**, *79*, 084505.

[4]     P. Monceau, *Adv. Phys.* **2012**, *61*, 325.

[5]     A. A. Balandin, R. K. Lake, T. T. Salguero, *Appl. Phys. Lett.* **2022**, *121*, 040401.

[6]     S. V Zaitsev-Zotov, *Physics-Uspekhi* **2004**, *47*, 533.

[7]     G. Grüner, *Rev. Mod. Phys.* **1988**, *60*, 1129.

[8]     A. A. Balandin, S. V Zaitsev-Zotov, G. Grüner, *Appl. Phys. Lett.* **2021**, *119*, 170401.

[9]     F. Kargar, Z. Barani, N. R. Sesing, T. T. Mai, T. Debnath, H. Zhang, Y. Liu, Y. Zhu, S. Ghosh, A. J. Biacchi, F. H. da Jornada, L. Bartels, T. Adel, A. R. H. Walker, A. V. Davydov, T. T. Salguero, R. K. Lake, A. A. Balandin, *Appl. Phys. Lett.* **2022**, *121*, 221901.

[10]    K. H. Choi, S. Oh, S. Chae, B. J. Jeong, B. J. Kim, J. Jeon, S. H. Lee, S. O. Yoon, C. Woo, X. Dong, A. Ghulam, C. Lim, Z. Liu, C. Wang, A. Junaid, J. H. Lee, H. K. Yu, J. Y. Choi, *J. Alloys Compd.* **2021**, *853*, 157375.

[11]    F. W. Boswell, A. Prodan, J. K. Brandon, *J. Phys. C Solid State Phys.* **1983**, *16*, 1067.

[12]    J. A. Galvis, A. Fang, D. Jiménez-Guerrero, J. Rojas-Castillo, J. Casas, O. Herrera, A. C. Garcia-Castro, E. Bousquet, I. R. Fisher, A. Kapitulnik, P. Giraldo-Gallo, *Phys. Rev. B* **2023**, *107*, 045120.

[13]    B. S. de Lima, N. Chaia, T. W. Grant, L. R. de Faria, J. C. Canova, F. S. de Oliveira, F. Abud, A. J. S. Machado, *Mater. Chem. Phys.* **2019**, *226*, 95.

[14]    X. Yang, Y. Zhou, M. Wang, H. Bai, X. Chen, C. An, Y. Zhou, Q. Chen, Y. Li, Z. Wang, J. Chen, C. Cao, Y. Li, Y. Zhou, Z. Yang, Z. A. Xu, *Sci. Rep.* **2018**, *8*, 1.

[15] R. Z. Xu, X. Du, J. S. Zhou, X. Gu, Q. Q. Zhang, Y. D. Li, W. X. Zhao, F. W. Zheng, M. Arita, K. Shimada, T. K. Kim, C. Cacho, Y. F. Guo, Z. K. Liu, Y. L. Chen, L. X. Yang, *npj Quantum Mater.* **2023**, *8*, 1.

[16] B. Guster, M. Pruneda, P. Ordejón, E. Canadell, *Phys. Rev. B* **2022**, *105*, 64107.

[17] Y. Yuan, W. Wang, Y. Zhou, X. Chen, C. Gu, C. An, Y. Zhou, B. Zhang, C. Chen, R. Zhang, Z. Yang, *Adv. Electron. Mater.* **2020**, *6*, 1901260.

[18] X. Zhang, Q. Gu, H. Sun, T. Luo, Y. Liu, Y. Chen, Z. Shao, Z. Zhang, S. Li, Y. Sun, Y. Li, X. Li, S. Xue, J. Ge, Y. Xing, R. Comin, Z. Zhu, P. Gao, B. Yan, J. Feng, M. Pan, J. Wang, *Phys. Rev. B* **2020**, *102*, 35125.

[19] H. Sun, Z. Shao, T. Luo, Q. Gu, Z. Zhang, S. Li, L. Liu, H. Gedeon, X. Zhang, Q. Bian, J. Feng, J. Wang, M. Pan, *New J. Phys.* **2020**, *22*, 083025.

[20] Y. Zhang, R. Zhou, H. Wu, J. S. Oh, S. Li, J. Huang, J. D. Denlinger, M. Hashimoto, D. Lu, S.-K. Mo, K. F. Kelly, R. J. Birgeneau, B. Lv, G. Li, M. Yi, *Phys. Rev. B* **2023**, *108*, 155121.

[21] S. Chae, A. Siddiqa, S. Oh, B. Kim, K. Choi, W.-S. Jang, Y.-M. Kim, H. Yu, J.-Y. Choi, *Nanomaterials* **2018**, *8*, 794.

[22] A. K. Geremew, S. Rumyantsev, M. A. Bloodgood, T. T. Salguero, A. A. Balandin, *Nanoscale* **2018**, *10*, 19749.

[23] D. W. Bullett, *J. Phys. C Solid State Phys.* **1984**, *17*, 253.

[24] Y. C. Luo, Y. Y. Lv, R. M. Zhang, L. Xu, Z. A. Zhu, S. H. Yao, J. Zhou, X. X. Xi, Y. B. Chen, Y. F. Chen, *Phys. Rev. B* **2021**, *103*, 64103.

[25] S. L. L. M. Ramos, B. R. Carvalho, R. L. Monteiro Lobato, J. Ribeiro-Soares, C. Fantini, H. B. Ribeiro, L. Molino, R. Plumadore, T. Heinz, A. Luican-Mayer, M. A. Pimenta, *ACS Nano* , DOI:10.1021/acsnano.3c03902.

[26] Y. Zhao, S. Han, J. Zhang, L. Tong, *J. Raman Spectrosc.* **2021**, *52*, 525.

[27] J. J. Gao, J. G. Si, X. Luo, J. Yan, F. C. Chen, G. T. Lin, L. Hu, R. R. Zhang, P. Tong, W. H. Song, X. B. Zhu, W. J. Lu, Y. P. Sun, *Phys. Rev. B* **2018**, *98*, 224104.

[28] A. C. Ferrari, A. A. Balandin, *Appl. Phys. Lett.* **2023**, *122*, 70401.

[29] F. Kargar, E. A. Coleman, S. Ghosh, J. Lee, M. J. Gomez, Y. Liu, A. S. Magana, Z. Barani, A. Mohammadzadeh, B. Debnath, R. B. Wilson, R. K. Lake, A. A. Balandin, *ACS Nano* **2020**, *14*, 2424.

[30] Y. Chen, P. Wang, M. Wu, J. Ma, S. Wen, X. Wu, G. Li, Y. Zhao, K. Wang, L. Zhang, L. Huang, W. Li, M. Huang, *Appl. Phys. Lett.* **2019**, *115*, 151905.

[31] J. C. Tsang, J. E. Smith, M. W. Shafer, *Phys. Rev. Lett.* **1976**, *37*, 1407.

[32] O. R. Albertini, R. Zhao, R. L. McCann, S. Feng, M. Terrones, J. K. Freericks, J. A. Robinson, A. Y. Liu, *Phys. Rev. B* **2016**, *93*, 214109.

[33] P. Goli, J. Khan, D. Wickramaratne, R. K. Lake, A. A. Balandin, *Nano Lett.* **2012**, *12*, 5941.

[34] R. Samnakay, D. Wickramaratne, T. R. Pope, R. K. Lake, T. T. Salguero, A. A. Balandin, *Nano Lett.* **2015**, *15*, 2965.

[35] E. M. Lacinska, M. Furman, J. Binder, I. Lutsyk, P. J. Kowalczyk, R. Stepniewski, A. Wysmolek, *Nano Lett.* **2022**, *22*, 2835.

[36] Y. Chen, P. Wang, M. Wu, J. Ma, S. Wen, X. Wu, G. Li, Y. Zhao, K. Wang, L. Zhang, L. Huang, W. Li, M. Huang, *Appl. Phys. Lett.* **2019**, *115*, 151905.

[37] F. Kargar, A. Krayev, M. Wurch, Y. Ghafouri, T. Debnath, D. Wickramaratne, T. T. Salguero, R. K. Lake, L. Bartels, A. A. Balandin, *Nanoscale* **2022**, *14*, 6133.

[38] Z. Barani, T. Geremew, M. Stokey, N. Sesing, M. Taheri, M. J. Hilfiker, F. Kargar, M. Schubert, T. T. Salguero, A. A. Balandin, *Adv. Mater.* **2023**, *35*, 2209708.

[39] E. Bjerkelund, A. Kjekshus, *J. Less-Common Met.* **1964**, *7*, 231.

[40] K. Selte, A. Kjekshus, C. S. Petersen, H. Halvarson, L. Nilsson, *Acta Chem. Scand.* **1964**, *18*, 690.

[41] J. P. Perdew, K. Burke, M. Ernzerhof, *Phys. Rev. Lett.* **1996**, *77*, 3865.

[42] P. E. Blochl, *Phys. Rev. B* **1994**, *50*, 17953.

[43] G. Kresse, D. Joubert, *Phys. Rev. B - Condens. Matter Mater. Phys.* **1999**, *59*, 1758.

[44] G. Kresse, J. Furthmüller, *Comput. Mater. Sci.* **1996**, *6*, 15.

[45] S. Grimme, J. Antony, S. Ehrlich, H. Krieg, *J. Chem. Phys.* **2010**, *132*, 154104.

[46] A. Togo, F. Oba, I. Tanaka, *Phys. Rev. B - Condens. Matter Mater. Phys.* **2008**, *78*, 134106.

[47] A. Togo, I. Tanaka, *Scr. Mater.* **2015**, *108*, 1.

[48] S. Zhu, W. Zheng, *J. Phys. Chem. Lett. 2021* **2022**, *12*, 57.

[49] A. A. Balandin, *Nat. Mater.* **2011**, *10*, 569.

[50] H. Malekpour, A. A. Balandin, *J. Raman Spectrosc.* **2018**, *49*, 106.

[51] S. Guo, Y. Xu, T. Hoke, G. Sohi, S. Li, X. Chen, *J. Appl. Phys.* **2023**, *133*, 120701.

[52] M. B. Walker, R. Morelli, *Phys. Rev. B* **1988**, *38*, 4836.

[53] S. Tadaki, N. Hino, T. Sambongi, K. Nomura, F. Lévy, *Synth. Met.* **1990**, *38*, 227.

[54] S. Yang, H. Cai, B. Chen, C. Ko, V. O. Özçelik, D. F. Ogletree, C. E. White, Y. Shen, S. Tongay, *Nanoscale* **2017**, *9*, 12288.

[55] S. Y. Chen, C. H. Naylor, T. Goldstein, A. T. C. Johnson, J. Yan, *ACS Nano* **2017**, *11*, 814.

[56] J. C. Bennett, F. W. Boswell, A. Prodan, J. M. Corbett, S. Ritchie, *J. Phys. Condens. Matter* **1991**, *3*, 6959.

[57] D. J. Eaglesham, D. Bird, R. L. Withers, J. W. Steeds, *J. Phys. C Solid State Phys.* **1985**, *18*, 1.

[58] T. Ikari, H. Berger, F. Levy, *Phys. status solidi* **1987**, *139*, K37.

[59] M. B. Walker, *Can. J. Phys.* **1985**, *63*, 46.

[60] F. W. Boswell, A. Prodan, J. C. Bennett, J. M. Corbett, L. G. Hiltz, *Phys. status solidi* **1987**, *102*, 207.

[61]  J. M. Corbett, L. G. Hiltz, F. W. Boswell, J. C. Bennett, A. Prodan, *Ultramicroscopy* **1988**, *26*, 43.

[62]  Z. Y. Chen, M. B. Walker, R. Morelli, *Phys. Rev. B* **1989**, *39*, 11742.

[63]  X. Luo, F. C. Chen, Q. L. Pei, J. J. Gao, J. Yan, W. J. Lu, P. Tong, Y. Y. Han, W. H. Song, Y. P. Sun, *Appl. Phys. Lett.* **2017**, *110*, 092401.

[64]  S. Reich, A. C. Ferrari, R. Arenal, A. Loiseau, I. Bello, J. Robertson, *Phys. Rev. B - Condens. Matter Mater. Phys.* , DOI:10.1103/PhysRevB.71.205201.

[65]  M. N. O. Sadiku, *Elements of electromagnetics*, Oxford University Press, New York, Oxford, Third., **2005**.